\documentclass[aps,preprint,epsfig,rotate]{revtex4}
\usepackage{graphicx}
\usepackage{bm}
\usepackage{epsfig}

\input epsf
\begin{document}
\title{Bound state spectra of three-body muonic molecular ions}

 \author{Alexei M. Frolov}
 \email[E--mail address: ]{afrolov@uwo.ca}

 \author{David M. Wardlaw}
 \email[E--mail address: ]{dwardlaw@uwo.ca}

\affiliation{Department of Chemistry\\
 University of Western Ontario, London, Ontario N6H 5B7, Canada}

\date{\today}

\begin{abstract}

The results of highly accurate calculations are presented for all twenty-two
known bound $S(L = 0)-, P(L = 1)-, D(L = 2)-$ and $F(L = 3)-$states in the
six three-body muonic molecular ions $pp\mu, pd\mu, pt\mu, dd\mu, dt\mu$ and
$tt\mu$. A number of bound state properties of these muonic molecular ions
have been determined numerically to high accuracy. The dependence of the
total energies of these muonic molecules upon particle masses is considered.
We also discuss the current status of muon-catalysis of nuclear fusion
reactions.

PACS number(s): 36.10.+Di, 36.10.-k and 31.10.+z.
\end{abstract}
\maketitle

\newpage

\section{Introduction}

In this study we consider the bound state spectra of the muonic molecular
ions $pp\mu, pd\mu, pt\mu, dd\mu, dt\mu$ and $tt\mu$. In this paper the
notations $p, d, t$ designate the nuclei of hydrogen isotopes (protium,
deuterium and tritium, respectively), while $\mu$ means the negatively
charged muon $\mu^{-}$. Our main goal is to determine total energies and
other bound state properties in these muonic molecular ions to high enough
numerical accuracy to be sufficient for all current and anticipated future
experimental needs.

In general, the bound state spectra in the six muonic molecular ions
$pp\mu, pd\mu, pt\mu, \linebreak dd\mu, dt\mu$ and $tt\mu$ can be separated
into three different groups \cite{Fro2001} on qualitative grounds. The first
group includes the three light muonic molecular ions $pp\mu, pd\mu$ and
$pt\mu$. Each of these systems has two bound states: one $S(L = 0)-$state
and one $P(L = 1)-$state, where the notation $L$ means the total angular
momentum of the three-body system. Neither of these two states is weakly
bound. Note that each of these light muonic molecular ions contains at least
one protium nucleus. The second group includes the two `intermediate' muonic
molecular ions $dd\mu$ and $dt\mu$ each of which has five bound states: two
$S(L = 0)-$states, two $P(L = 1)-$states and one $D(L = 2)-$state. One of
these five states (the excited $P^{*}(L = 1)-$state) in each of these ions
is weakly bound. The third group contains only the heaviest muonic molecular
ion $tt\mu$ which has six bound states (and no weakly bound states): two
$S(L = 0)-$states, two $P(L = 1)-$states, one $D(L = 2)-$state and one $F(L
= 3)-$state.

It can be shown (see, e.g., \cite{FrBi}) that the total number of bound
states in any muonic molecular ion $a^{+} b^{+} \mu^{-}$ is determined by
the lightest nucleus in this ion. This explains why only three groups of
different bound state spectra can be found in six such ions. Moreover, it
follows that there must be a similarity between the energy spectra of the
`protium' muonic molecular ions $pp\mu, pd\mu$ and $pt\mu$. An analogous
similarity can be found in the bound state spectra of the $dd\mu$ and
$dt\mu$ ions. It can be shown that in such `families' of muonic molecular
ions the symmetric ion, e.g., $pp\mu$, always has the maximal binding energy
of the three protium ions: $pp\mu, pd\mu$ and $pt\mu$. By using these
similarities between the bound state spectra in each of these `families',
one also finds a number of useful relations for the total and binding
energies as well as for other bound state properties of different muonic
molecular ions. For instance, let us assume that we know that the excited
$P-$state in the $dd\mu$ ion is weakly bound and its binding energy is
$\approx$ -1.9745 $eV$ (see, e.g., \cite{Fro2001}). From the similarity of
the bound state spectra of the $dd\mu$ and $dt\mu$ ions it one predicts that
the corresponding excited $P-$state in the $dt\mu$ ion is also weakly bound
and its binding energy is above -1.9745 $eV$, i.e. -1.9745 $eV$ $\le
\varepsilon(dt\mu, P^{*}(L = 1)) < 0$. This prediction turns out to be
correct, as the $dt\mu$ ion has a weakly bound excited $P-$state with the
binding energy $\approx$ -0.660 $eV$.

Our labels for the bound states in muonic molecular ions are based on atomic
$LS$-notations (see, e.g., \cite{Sob}). Note that there is another
classification scheme which is still in use for muonic molecular ions and
which was originally introduced to classify bound state spectra in adiabatic
molecular ions, e.g., in the H$^{+}_{2}$ molecular ion \cite{BoOp}. In this
scheme each bound state is designated by its rotational $J$ and vibrational
$\nu$ quantum numbers, i.e., we have the $(J, \nu)$-states. The ground state
in any muonic molecular ion is designated as the (0,0)-state, while the
excited $P-$state in this scheme is denoted as the (1,1)-state, etc. Each of
these classification schemes has its own advantages and disadvantages in
applications to actual systems. Note also that there is a uniform
correspondence between the `atomic' and `molecular' classification schemes.

\section{The Hamiltonian and wave functions}

As mentioned above in this study we consider the bound state spectra in the
six muonic molecular ions $pp\mu, pd\mu, pt\mu, dd\mu, dt\mu$ and $tt\mu$.
All particles which form such three-body ions are assumed to be point and
structureless. Each of these three particles has a finite mass which equals
one of the masses $m_{\mu}, m_p, m_d$ and/or $m_t$; the electric charges are
$q_{\mu} = -1$ and $q_p = q_d = q_t = +1$ (in muon-atomic units, where
$\hbar = 1, m_{\mu} = 1, e = 1$). The Hamiltonian $H$ of the three-body
muonic molecular ions is written in the form, e.g., for the $a^{+}b^{+}
\mu^{-}$ ion
\begin{equation}
 H = -\frac{\hbar^{2}}{2 m_{\mu}} \Bigl( \frac{m_{\mu}}{m_a} \nabla^{2}_{a}
 + \frac{m_{\mu}}{m_b} \nabla^{2}_{b} + \nabla^{2}_{\mu} \Bigr) +
 \frac{q_a q_b}{r_{ab}} + \frac{q_a q_{\mu}}{r_{a{\mu}}} + \frac{q_b
 q_{\mu}}{r_{b{\mu}}} \label{Ham}
\end{equation}
where $\nabla_{i} = \Bigl( \frac{\partial}{\partial x_{i}},
\frac{\partial}{\partial y_{i}}, \frac{\partial}{\partial z_{i}} \Bigr)$ and
$i = a, b, \mu$. In muon-atomic units we have
\begin{equation}
 H = -\frac12 \Bigl( \frac{1}{m_a} \nabla^{2}_{a} + \frac{1}{m_b}
 \nabla^{2}_{b} + \nabla^{2}_{\mu} \Bigr) + \frac{1}{r_{ab}} -
 \frac{1}{r_{a{\mu}}} - \frac{1}{r_{b{\mu}}} \label{Ham1}
\end{equation}
where the two masses $m_a$ and $m_b$ of nuclei of two hydrogen isotopes must
be expressed in terms of the muon mass $m_{\mu}$. In fact, in this study
only the muon-atomic units ($\hbar = 1, m_{\mu} = 1, e = 1$) are used.
Advantages of these units are discussed in Section IV below.

Our computational goal is to determine exceptionally accurate solutions,
i.e., the eigenstates and corresponding wave functions of the
non-relativistic Schr\"odinger equation $H \Psi({\bf r}_1, {\bf r}_2, {\bf
r}_3) = E \cdot \Psi({\bf r}_1, {\bf r}_2, {\bf r}_3)$, where $E < 0$ and
the non-relativistic Hamiltonian is written in the form of Eq.(\ref{Ham1}).
In actual calculations the wave functions of muonic molecular ions are
usually approximated with the use of different variational expansions. In
this work we shall consider the exponential variational expansion in
relative coordinates $r_{12}, r_{13}, r_{23}$ \cite{Fro01}. Here and
everywhere below in this work the notation $r_{ij} = \mid {\bf r}_i - {\bf
r}_j \mid = r_{ji}$ designates the relative coordinate between particles $i$
and $j$. In many cases, however, it is very convenient to introduce three
new variables $u_1, u_2, u_3$ which are called perimetric coordinates. They
are simply related to the three relative coordinates: $u_i = \frac12 (r_{ik}
+ r_{ij} - r_{jk})$, and therefore, $r_{ij} = u_i + u_j$, where $(i, j, k) =
(1, 2, 3)$. The perimetric coordinates are truly independent, and each of
them varies from 0 to $+ \infty$. This significantly simplifies derivation
of the explicit formulas for all matrix elements needed in highly accurate
computations of the bound states. The explicit form of the exponential
variational expansion in perimetric/relative coordinates is
\begin{eqnarray}
 \Psi_{LM} = \frac{1}{2} (1 + \kappa \hat{P}_{21}) \sum_{i=1}^{N}
 \sum_{\ell_{1}} C_{i} {\cal Y}_{LM}^{\ell_{1},\ell_{2}} ({\bf r}_{31},
 {\bf r}_{32}) \phi_i(r_{32},r_{31},r_{21}) \exp(-\alpha_{i} u_1 -
 \beta_{i} u_2 - \gamma_{i} u_3) \times \label{e1} \\
 \exp(\imath \delta_{i} u_1 + \imath e_{i} u_2
      + \imath f_{i} u_3) \; \; \; , \nonumber
\end{eqnarray}
where $C_{i}$ are the linear (or variational) parameters, $\alpha_i,
\beta_i, \gamma_i, \delta_i, e_i$ and $f_{i}$ are the non-linear parameters
and $\imath$ is the imaginary unit.

The functions ${\cal Y}_{LM}^{\ell_{1},\ell_{2}} ({\bf r}_{31}, {\bf
r}_{32})$ in Eq.(\ref{e1}) are the bipolar harmonics \cite{Varsh} of the two
vectors ${\bf r}_{31} = r_{31} \cdot {\bf n}_{31}$ and ${\bf r}_{32} =
r_{32} \cdot {\bf n}_{32}$. The bipolar harmonics are defined as follows
\cite{Varsh}
\begin{equation}
 {\cal Y}_{LM}^{\ell_{1},\ell_{2}} ({\bf x}, {\bf y}) = x^{\ell_{1}}
 y^{\ell_{2}} \sum_{\ell_{1}, \ell_{2}} C^{LM}_{\ell_{1} m_{1};\ell_{2}
 m_{2}} Y_{\ell_{1} m_{1}} ({\bf n}_{x}) Y_{\ell_{2} m_{2}} ({\bf n}_{y})
\end{equation}
where $C^{LM}_{\ell_{1} m_{1};\ell_{2} m_{2}}$ are the Clebsch-Gordan
coefficients (see, e.g., \cite{Varsh}) and the vectors ${\bf n}_{x} =
\frac{{\bf x}}{x}$ and ${\bf n}_{y} = \frac{{\bf y}}{y}$ are the
corresponding unit vectors constructed for arbitrary non-zero vectors ${\bf
x}$ and ${\bf y}$. Also, in this equation $L$ is the total angular momentum
of the three-body system, i.e. $\hat{L}^2 \Psi_{LM} = L (L + 1) \Psi_{LM}$,
while $M$ is the eigenvalue of the $\hat{L}_z$ operator, i.e. $\hat{L}_z
\Psi_{LM} = M \Psi_{LM}$. In actual calculations it is possible to use only
those bipolar harmonics for which $\ell_{1} + \ell_{2} = L + \epsilon$,
where $\epsilon = 0$ or 1. The first choice of $\epsilon$ (i.e. $\epsilon =
0$) corresponds to the natural spatial parity $\chi_P = (-1)^L$ of the wave
functions. The second choice (i.e. $\epsilon = 1$) represents states with
the unnatural spatial parity $\chi_P = (-1)^{L+1}$. In this work we shall
consider only the bound states of natural parity, since only such states
exist in real physical systems. An additional family of polynomial-type
functions $\phi_i(r_{32}, r_{31}, r_{21})$ are also used in Eq.(\ref{e1}) to
represent the inter-particle correlations at short distances. In general,
these simple polynomial functions allow one to increase the overall
flexibility of the variational expansion Eq.(\ref{e1}). In our present
calculations, however, these additional functions were chosen in the form
$\phi_i(r_{32}, r_{31}, r_{21}) = 1$ for $i = 1, \ldots, N$. The operator
$\hat{P}_{21}$ in Eq.(\ref{e1}) is the permutation of the identical
particles in symmetric three-body systems, where $\kappa = \pm 1$, otherwise
$\kappa = 0$. In this study we assume that $\kappa = \chi_P = (-1)^L$ for
all symmetric muonic molecular ions $pp\mu, dd\mu, tt\mu$ and $\kappa = 0$
for all non-symmetric ions $pd\mu, pt\mu, dt\mu$.

In general, highly accurate computations of bound states in muonic molecular
ions are not easy to perform, since there are bound states with different
angular momenta $L$ ($L$ = 0, 1, 2 and 3) and some of these states are very
weakly bound. Variational expansions used in highly accurate calculations
must provide fast convergence rates for each of the bound states, including
all weakly bound states. Note that the actual goal of many current
calculations of muonic molecular ions is the computation of various bound
state properties, rather than the energies. In general, the convergence rate
for some of these properties, including many nuclear-nuclear expectation
values, e.g., the expectation values which include the nuclear-nuclear
delta-function $\delta_{++}$ (e.g., the $\langle \delta_{++}
\frac{\partial^n}{\partial r^n_{++}} \rangle$ expectation values for $n \ge
1$), is substantially slower than for the total energies. This explains our
current need for highly accurate wave functions. For instance, as follows
from numerical calculations to determine the $\langle \delta_{++}
\frac{\partial^n}{\partial r^n_{++}} \rangle$ expectation values, to the
accuracy $\pm 1 \cdot 10^{-8}$ one needs to use wave functions which provide
an accuracy $\approx 1 \cdot 10^{-15}$ $a.u.$ for the total energy. Such
values are needed in computations of the lowest order relativistic and QED
corrections. A separate, but serious problem for accurate computations of
three-body systems is the adiabatic divergence described in \cite{JETP}.
This problem always appears when `pure atomic' variational expansions are
applied to the two-center Coulomb systems and/or to the systems close to
them. For our systems this means that the convergence rates observed for
the $pp\mu$ and $pd\mu$ ions are relatively high in comparison to the
analogous convergence rates for the $dt\mu$ and $tt\mu$ ions which are
substantially slower.

Nevertheless, variational calculations of bound states in muonic molecular
ions are of great interest for predicting the physics of few-body systems
as well as in some applications. In general, the study of bound state
spectra in muonic molecular ions has provided us with a large amount of
very valuable information and drastically improved our knowledge about the
bound state spectra in arbitrary Coulomb three-body systems. Furthermore,
all muonic molecular ions are three-body systems with unit charges. The
energy spectra in such systems have many significant differences from known
atomic spectra. In particular, any Coulomb three-body system with unit
charges has a finite number of bound states \cite{Fro1999}. The only
exception to this rule is the ${}^{\infty}$H$^{+}_2$ ion which has infinite
number of bound states \cite{Fro1999}.

\section{Optimization of the non-linear parameters in trial wave functions}

Optimization of the non-linear parameters in trial wave functions is a
central part of construction of any highly accurate solution of the
Schr\"{o}dinger equation. By optimizing the non-linear parameters in some
rapidly convergent variational expansions one can produce extremely
accurate and compact wave functions for arbitrary three-body systems. Such
wave functions can be used in further calculations of various bound state
properties and different corrections to the non-relativistic energies of
three-body systems. In this study we construct our highly accurate
variational wave functions, Eq.(\ref{e1}), with the use of a two-stage
optimization strategy of the non-linear parameters \cite{Fro01}. The
two-stage procedure consists of: (a) construction of a short-term booster
wave function with carefully optimized non-linear parameters, and (b)
quasi-random choice of the remaining non-linear parameters from a few
(usually, three - five) optimal boxes (or parallelotops). To simplify the
description of our optimization procedure below we shall assume that all
exponents in Eq.(\ref{e1}) are real. In other words, all parameters
$\delta_i, e_i$ and $f_i$ in Eq.(\ref{e1}) equal zero identically for $i =
1, \ldots, N$. In fact, these later parameters are really needed only for
highly accurate calculations of the pure adiabatic three-body systems and
systems close to them. Among all muonic molecular ions $pp\mu, pd\mu, pt\mu,
dd\mu, dt\mu$ and $tt\mu$ even the two heaviest systems $dt\mu$ and $tt\mu$
are not close to adiabatic systems, examples of which are the one-electron
DT$^{+}$ and T$_2^{+}$ ions.

For muonic molecular ions considered in this study the short-term booster
wave function, Eq.(\ref{e1}), usually includes $N_0$ = 400 - 600 basis
functions (exponents) with $3 N_0$ = 1200 - 1800 non-linear parameters in
them. All these parameters must carefully be optimized. After such an
optimization the short-term booster function provides 11 - 15 exact decimal
digits for each bound state energy in the considered muonic molecular ions.
It appears that the overall accuracy of such short-term wave functions is
much better for the protium muonic molecular ions $pp\mu, pd\mu, pt\mu$ than
for the heavier ions $tt\mu$ and $dt\mu$. At the second stage of our
optimization procedure the remaining $3 (N - N_0)$ non-linear parameters in
the wave function Eq.(\ref{e1}) are chosen quasi-randomly from three
different boxes or parallelotops. The total energies and other bound state
properties obtained with such trial wave functions depend upon the
boundaries of these boxes. In reality the boundaries of these three boxes
can be described \cite{Fro01} with the use of 28 non-linear parameters only.
The numerical values of these 28 parameters were optimized approximately
with the use of $N$ = 800, 1000, 1200 and 1400 basis functions \cite{Fro01}.
These values allowed us to determine the approximate limit for each of these
28 parameters as $N \rightarrow \infty$ by extrapolation. These limiting
values have been used in our final computations.

The two-stage strategy proposed in \cite{Fro2001} and described above allows
one to obtain very accurate variational wave functions based on the use of
exponents in relative and/or perimetric coordinates. In particular, the
overall accuracy of our results obtained in this study for all considered
muonic molecular ions (see Tables I, II, III and IV below) is significantly
higher than the accuracy obtained for these ions in earlier calculations.
Moreover, by using this optimization strategy we expect to be able to
increase the accuracy in future calculations by a factor of $\approx 10^3 -
10^5$ which would be sufficient for all future anticipated theoretical
needs. The described two-stage optimization procedure has been used in this
study for all bound $S(L = 0)-$ and $P(L = 1)-$states in muonic molecular
ions, including all excited states. For the bound $D(L = 2)-$states in heavy
muonic molecular ions ($dd\mu, dt\mu$ and $tt\mu$) we used another approach
in which the short-term booster function is not constructed. However, it is
clear that our current strategy for optimization of the non-linear
parameters in the wave functions of the bound $D(L = 2)-$states is not
optimal. In future studies we want to improve this strategy and produce
highly accurate results for all bound $D(L = 2)-$states in muonic molecular
ions.

\section{Variational energies of muonic molecular ions}

The results of our calculations of different bound states can be found in
Tables I - VI. All computations are performed in muon-atomic units, where
$m_{\mu} = 1, \hbar = 1$ and $e = 1$ and we used the following values of
nuclear masses \cite{COD}, \cite{CRC}:
\begin{eqnarray}
 m_{\mu} = 206.768262 m_e \; \; \; , \; \; \; m_p = 1836.152701 m_e \\
 m_{d} = 3670.483014 m_e \; \; \; , \; \; \; m_t = 5496.92158 m_e \nonumber
\end{eqnarray}
where $m_e$ designates the electron mass. Note that our highly accurate
computations in this study are performed with the use of 84 - 104 decimal
digits per computer word \cite{Bail1}, \cite{Bail2}, allowing total energies
to be determined to an accuracy $\approx 1 \cdot 10^{-20} - 1 \cdot
10^{-23}$ $m.a.u.$ A natural and effective way to perform such calculations
is to assume that all particle masses and corresponding conversion factors
(e.g., the factor $Ry$ below) are exact. Such assumptions are always made in
papers on highly accurate computations in few-body systems (see, e.g.,
\cite{Hili} and \cite{BaFr}). The known experimental uncertainties in
particle masses and conversion factors are taken into account at the last
step of calculations, when the most accurate computations are simply
repeated for a few times with the use of different particle masses and
conversion factors. Analogously, the lowest order relativistic and QED
corrections can be determined as the expectation values of some operators
computed with our non-relativistic wave functions. To avoid a substantial
loss of numerical accuracy during such computations these non-relativistic
wave functions must be extremely accurate.

Table I contains the total variational energies obtained for the ground $S(L
= 0)-$states of the non-symmetric muonic molecular ions $pd\mu, pt\mu$ and
$dt\mu$. Table II includes the total energies for the excited $S^{*}(L =
0)-$states of the $dd\mu, tt\mu$ and $dt\mu$ muonic molecular ions. The
total energies of the ground $S(L = 0)-$ and `rotationally' excited $P(L =
1)-$states of the symmetric muonic molecular ions $pp\mu, dd\mu$ and $tt\mu$
are presented in Table III - IV. Table V contains the best variational
energies obtained in our computations of the $D(L = 2)-$states of the
$dd\mu, tt\mu$ and $dt\mu$ muonic molecular ions. Highly accurate
computations of these bound $D(L = 2)-$states have been performed with the
use of extended arithmetic (for discussion and references, see,
\cite{BaFr}). In earlier works the bound $D(L = 2)-$states have been
determined with the use of quadruple precision only. However, our current
total energies determined for the bound $D(L = 2)-$states are not as
accurate as the total energies obtained for the bound $S(L = 0)-$ and $P(L =
1)-$states (see comment above).

Note also that the total energies of the $P(L = 1)-$states of the
non-symmetric muonic molecular ions $pd\mu, pt\mu$ and $dt\mu$ and excited
$P^{*}(L = 1)-$states of the $dd\mu, tt\mu$ and $dt\mu$ ions have recently
been determined in \cite{FrWa2010}. The most accurate variational energies
for all known 22 bound states in the set of six muonic molecular ions
studied here (expressed in muon-atomic units) can be found in Table VI. Note
that the $F(L = 3)-$state of the $tt\mu$ ion has not been re-calculated in
this study, but instead it has been taken from our earlier work
\cite{Fro2001}, where this state was computed with the use of quadruple
precision.

As follows from Tables I - VI the total energies obtained in this study for
different bound states in six muonic molecular ions are significantly more
accurate than the corresponding energies computed in earlier studies (see,
e.g., \cite{Fro2001}, \cite{Hili} and \cite{BaFr}). The current wave
functions are more compact and have better overall quality than wave
functions obtained in \cite{Fro2001}, \cite{Hili} and \cite{BaFr}. They can
be used for highly accurate computations of other bound state properties,
including properties which contain singular expectation values.

Our variational wave functions can be used to compute some bound state
properties of muonic molecular ions. A large number of bound state
properties have been computed in our earlier studies (see, e.g.,
\cite{Fro01} and references therein). However, some of the bound state
properties could not be determined to high numerical accuracy, due to
relatively low accuracy of the wave functions used in earlier studies. It
was clear that the expectation values of some nuclear-nuclear properties,
e.g., all properties which include the nuclear-nuclear delta-function,
needed to be re-calculated with more accurate wave functions. By using
highly accurate wave functions obtained in this work we have performed
numerical re-calculation of a number of bound state properties for different
muonic molecular ions.

The computed expectation values can be found in Tables VII and VIII. Results
presented in Table VII illustrate convergence of some expectation values
upon the total number $N$ of basis functions used. By comparing the
expectation values computed with the use of different number(s) of basis
functions we have determined the corresponding asymptotic values for $N
\rightarrow \infty$. Formally, this procedure allows one to determine the
number of stable decimal digits for each of the computed expectation values.
Table VIII contains some bound state properties determined for the ground
$S(L = 0)-$ states in the $pd\mu$ and $dt\mu$ muonic molecular ions and for
the excited $S(L = 0)-$state in the $dt\mu$ ion.

\section{Mass dependence of the total energies for muonic molecular ions}

Let us consider the problem of mass dependence for the total energies of
muonic molecular ions. Briefly, this problem is formulated as follows. The
Hamiltonian $H$ of any muonic molecular ion, Eq.(\ref{Ham}), contains three
different masses $m_a, m_b, m_{\mu}$. These particle masses are the subject
of constant experimental revision. Furthermore, various authors often use
slightly different particle masses in their calculations. For highly
accurate computations this means the almost constant necessity of
re-calculation of the energies and corresponding wave functions. In fact,
all masses of particles which form muonic molecular ions are currently known
to relatively high experimental accuracy, i.e. all possible
`mass-corrections' must be very small. This means that such corrections can
be considered by using various methods of perturbation theory. Nevertheless,
it is very interesting in some  cases to evaluate corrections produced by
the corresponding `mass shift'. The most important bound state property is
the total energy $E$. The first order variation of $E$ with the particle
masses is written in the form
\begin{equation}
 E_{new} = E_{our} + \alpha \Bigl[ \frac{m_{\mu}(new)}{m_{t}(new)} -
 \frac{m_{\mu}(our)}{m_{t}(our)} \Bigr] + \beta \Bigl[
 \frac{m_{\mu}(new)}{m_{d}(new)} - \frac{m_{\mu}(our)}{m_{d}(our)} \Bigr]
 \label{eq3}
\end{equation}
in the case of the $dt\mu$ ion. In Eq.(\ref{eq3}) and Eq.(\ref{eq4}) below
all energies must be expressed in the same units, e.g., in atomic units or
in muon-atomic units reduced to the same muon mass. The same formula can be
written for any non-symmetric muonic molecular ion. In such cases we always
have $\alpha \geq \beta$, if the coefficient $\alpha$ corresponds to the
mass shift produced by the heaviest nucleus. For symmetric muonic molecular
ions, e.g., for the $dd\mu$ ion, the analogous formula is
\begin{equation}
 E_{new} = E_{our} + \alpha \Bigl[ \frac{m_{\mu}(new)}{m_{d}(new)} -
 \frac{m_{\mu}(our)}{m_{d}(our)} \Bigr] \label{eq4}
\end{equation}
In these equations the notation 'our' denotes the mass value used in this
study, while the notation 'new' designates a different mass value, e.g. from
some work performed in the future. In general, the numerical values of the
coefficients $\alpha$ and $\beta$ in Eq.(\ref{eq3}) (also called the mass
gradients) are determined from separate energy calculations with different
masses. In our earlier works we have used a very simple approach based on
four additional calculations with different masses for non-symmetric muonic
molecular ions \cite{FSB}. For symmetric muonic molecular ions one has to
perform at least two additional calculations with the two different mass
ratios. This method is simple, but it is not very accurate.

Recently, we have developed a more accurate procedure. To describe this
procedure let us consider the ground $P(L = 1)-$state in the $dd\mu$ ion.
The mass ratio $\frac{m_{\mu}(our)}{m_{d}(our)}$ is designated below as
$x_0$, while the notation $h$ stands for the difference
$\frac{m_{\mu}(new)}{m_{d}(new)} - \frac{m_{\mu}(our)}{m_{d}(our)}$ used in
Eq.(\ref{eq4}). In these notations the total energies $E_{our}$ and
$E_{new}$ are $E_{our} = E(x_0)$ and $E_{new} = E(x_0 + h)$, respectively.
The absolute value of $h$ is assumed to be small. Our new method for
calculation of the mass gradient $\alpha$ is based on calculation of the
four ground state energies $E(x_0 - 2 h), E(x_0 - h), E(x_0 + h) = E_{new}$
and $E(x_0 + 2 h)$. These four energies are determined in calculations
performed with the use of the maximal number of basis functions (at the
maximal dimension). No re-optimization of non-linear parameters in the wave
function is required during such additional computations. The `mass
gradient' $\alpha$ in Eq.(\ref{eq4}) can now be computed with the use of the
formula
\begin{equation}
 \alpha = \frac{1}{12 h} \Bigl[ E(x_0 - 2 h) - 8 E(x_0 - h) + 8 E(x_0 + h)
 - E(x_0 + 2 h) \Bigr] + \frac{h^4}{30} A \label{massgr}
\end{equation}
where $A$ is a numerical parameter of order of unity. It can be shown that
this parameter equals the fifth order derivative of the total energy $E$
with respect to the mass ratio $\frac{m_{\mu}}{m_{d}}$ computed in some
point between $x_0 - 2 h$ and $x_0 + 2 h$. The formula, Eq.(\ref{massgr}),
has a very good numerical accuracy in actual applications to all bound
states, except weakly-bound states (discussed below).

The results of numerical calculations of the mass gradients $\alpha$ in
Eq.(\ref{massgr}) for the $P(L = 1)-$states in the symmetric $pp\mu, dd\mu$
and $tt\mu$ muonic molecular ions are presented in Table IX. Table IX also
contains the results for the excited $P^{*}(L = 1)-$state in the $tt\mu$
ion. All these energies have been re-calculated to the `standard' muon mass
$m_{\mu}$ = 206.768262 $m_e$. This Table also contains all intermediate
energies $E(x_0 - 2 h), E(x_0 - h), E(x_0 + h)$ and $E(x_0 + h)$ needed in
such calculations. As follows from Eq.(\ref{massgr}) our mass gradients
$\alpha$ are accurate to 11 - 13 decimal digits. The same procedure can be
used for other bound states in muonic molecular ions.

\subsection{Mass dependence for the weakly-bound states}

The mass dependence of the total and binding energies for weakly bound
states cannot be investigated with the use of the method described above,
since the total energies $E(x_0 + h)$ and $E(x_0 + 2 h)$ needed in
Eq.(\ref{massgr}) can correspond to unbound states. This indicates a
necessity to study the mass dependence of the total and binding energies for
weakly bound states. Formally, to answer this question one needs to perform
a number of calculations with different particle masses. For simplicity, let
us discuss the weakly bound excited $P^{*}(L = 1)-$state in the $dt\mu$ ion.
If the muon mass $m_{\mu}$ and the muon-deuterium mass ratio $\tau =
\frac{m_{\mu}}{m_d}$ increase, then the $P^{*}(L = 1)-$state in the $dt\mu$
ion will be less and less bound. Finally, at some critical values of
$m_{\mu}$ and $\tau$ the $P^{*}(L = 1)-$state in the $dt\mu$ ion will become
unbound.

The approach based on the use of `conventional wisdom' allows one quickly to
find the numerical value of the critical muonic mass. Indeed, the
Hamiltonian of muonic molecular ions, Eq.(\ref{Ham1}), is a linear function
of the muonic mass $m_{\mu}$ and/or the mass ratio $\tau =
\frac{m_{\mu}}{m_d}$. Therefore, by performing calculations of the total
energy of some bound state for different muonic masses $m_{\mu}$ one finds
the dependence $E(m_{\mu})$, or $E(m_{\tau})$. The next step is to determine
the critical mass for which $E(m_{\mu}) = -\frac12 \frac{m_{\mu}
m_t}{m_{\mu} + m_t}$, i.e. the binding energy of this state equals zero. For
a very short interval of variation of $m_{\mu}$ the function $E(m_{\mu})$ is
almost a linear function upon the muonic mass $m_{\mu}$, then it is
relatively easy to obtain an approximate value of the critical muonic mass.

This approach was used in some earlier works, but it produces incorrect
results in applications to the bound states which can disappear (as bound
states) during variations of some physical parameter, e.g., the mass of the
particle. For such states one needs to use an alternative approach described
in \cite{Baz}. In this method the explicit expression for the total energy
of the two-body `almost unbound' system takes the form
\begin{equation}
 \varepsilon = - \frac{\pi^2}{16} \frac{(\mid U \mid - \mid U_1
 \mid)^2}{\mid U_1 \mid} = - \frac{\pi^2 \hbar^2}{32} V^2 \frac{(m_1 -
 m)^2}{m^2 m^2_{1}} \label{Baz}
\end{equation}
where $U = \frac{\hbar^2}{2 m} V$ is the effective two-body potential in the
two-body $d\mu + d$ system. The value $U_1$ corresponds to the mass $m_1$ at
which $\varepsilon = 0$. The difference of masses $m_1$ and $m$ is assumed
to be small in comparison with each of the two masses ($m_1$ and $m$). The
formula, Eq.(\ref{Baz}), predicts a quadratic dependence of the energy
$\varepsilon$ upon the mass difference $\Delta = m_1 - m$, i.e.
$\varepsilon(m_1, m) = \varepsilon(m_1, \Delta) \approx a \cdot
\frac{\Delta^2}{m^4_1}$. Although the actual $\varepsilon(m_1, \Delta)$
dependence must be described by the formula Eq.(\ref{Baz}) to good accuracy,
in reality however it is difficult to obtain the quadratic dependence for
the $\varepsilon(m_1, \Delta)$ function (on the parameter $\Delta$) by using
the results of numerical calculations with the same number of basis
functions. It is clear that in accurate calculations of bound states with
$\Delta \rightarrow 0$ one needs to increase the total number of basis
functions and constantly re-optimize the non-linear parameters of the
method. If such conditions are obeyed, then we can observe the expected
quadratic dependence of the binding energy $\varepsilon$ on $\Delta$.
Otherwise, from the results of such calculations one can see only an
approximate linear dependence.

Note that for real muonic molecular ions the total energy of the two-body
system $\varepsilon$ given in Eq.(\ref{Baz}) is, in fact, the binding energy
of the three-body ion, e.g., the $dt\mu$ ion, which corresponds to the
lowest-by-energy decay channel: $(dt\mu)^{+} = t\mu$(ground state) +
$d^{+}$. For the $dt\mu$ ion the mass $m$ in Eq.(\ref{Baz}) is the `reduced'
mass $m = \frac{(m_t + m_{\mu}) m_d}{m_t + m_{\mu} + m_d}$, while $m_1 =
\frac{(m_t + \tilde{m}_{\mu}) m_d}{m_t + \tilde{m}_{\mu} + m_d}$ is the
`threshold' mass, i.e. the reduced muonic mass for which the binding energy
equals zero.

The results of our calculations for the $P^{*}(L = 1)-$state in the $dd\mu$
ion can be found in Table X. In these calculations we have increased (at
each step) the mass of the $\mu^{-}$ muon (see above) by one electron mass
$m_e$. As follows from Table X the binding energy of the $P^{*}(L =
1)-$state in the $dd\mu$ ion decreases to zero. The value of muon mass at
which the corresponding bound state becomes unbounded is called the
threshold mass. For the $dd\mu$ ion such a muon threshold mass is designated
as $\tilde{m}_{\mu}$. In calculations for Table X we have used $N$ = 3300
basis functions. The $\varepsilon(m_1, \Delta)$ dependence which follows
from the results of these calculations can be represented as a linear
function. The threshold value of $\tilde{m}_{\mu}$ found from this linear
dependence approximately equals to the value $\tilde{m}_{\mu} = m_{\mu} +
7.95 m_e \approx 214.718262 m_e$. As follows from Table X the binding
energy of the $P^{*}(L = 1)-$state in the $dd\tilde{\mu}$ ion is $\approx$
-0.00002551843452 $eV$. This energy was obtained with $N$ = 3300 basis
functions. However, if we take 3840 basis functions, then the binding energy
of the $P^{*}(L = 1)-$state in the same $dd\tilde{\mu}$ ion is $\approx$
-0.00007865184962 $eV$. i.e., it corresponds to the state which is slightly
better bound. The muon threshold mass is always shifted to larger values of
$\tilde{m}_{\mu}$ if the total number of basis functions increases. This
example explains the occurence of what is often called ``running'' mass
threshold, a non-physical effect known to exist in numerical calculations of
any weakly-bound state. It is interesting to note that the muon mass
$m_{\mu} = 214.718262 m_e$ exceeds our `standard' muon mass (206.768262
$m_e$) already by $\approx$ 4 \%, but the $P^{*}(L = 1)-$state in the
$dd\mu$ ion is still bound (earlier estimations for such a mass deviation
were around 0.5 - 1 \%).

Analogous results for the weakly bound $P^{*}(L = 1)-$state in the $dt\mu$
ion can be found in Table XI. In computations performed for this Table we
varied only the muon mass $m_{\mu}$, while the deuterium and tritium masses
have not been changed. At such conditions the muon threshold mass was found
to be equal $\tilde{m}_{\mu} \approx m_{\mu} + 1.99 m_e$. It is clear that
the binding energy of the $P^{*}(L = 1)-$state in the $dt\mu$ ion is the
function of the two independent mass ratios $\frac{m_{\mu}}{m_d}$ and
$\frac{m_{\mu}}{m_t}$. Therefore, Table XI gives only an approximate picture
of how the total and binding energies (in eV) of the $P^{*}(L = 1)-$state in
the $dt\mu$ ion vary when the muon mass changes. To study the pre-threshold
mass dependence of the weakly bound $P^{*}(L = 1)-$state in the $dt\mu$ ion
in detail one also needs to consider changes of the deuterium and tritium
masses. Such calculations, however, are very difficult to perform, since
they require substantial computer resourses.

\section{Muon sticking probabilities}

Originally, all numerical computations of bound states in muonic molecular
ions were motivated by various problems of muon-catalyzed nuclear fusion. In
fact, the bound state computation of muonic molecular ions is only one of
many problems which must be solved before we can discuss a possibility to
use muon-catalyzed nuclear fusion for energy production and for other
purposes. It is clear that the most interesting and promising case is the
muon catalysis of the $(d,t)-$nuclear reaction in the $dt\mu$ ion. A central
problem here is to determine the muon sticking probability during the
nuclear reaction $dt\mu = {}^4$He$ + \mu + n$ since the numerical value of
this coefficient essentially determines the feasibility of using muon
catalysis of nuclear fusion reactions for energy production purposes. Let us
evaluate the muon sticking probabilities for this ion by assuming that the
nuclear $dt$-fusion occurs only in the two $S(L = 0)-$states (ground and
excited), ignoring the possibility of nuclear fusion in the $P(L = 1),
P^{*}(L = 1)$ and $D(L = 2)$ bound states of the $dt\mu$ ion.

The analytical expression for the muon sticking probability for the bound
$S(L = 0)-$state (initial state is designated with the subscript $in$; final
state by $fi$) takes the form \cite{Fro01} (see also \cite{Hu1} and
\cite{Hu2})
\begin{eqnarray}
 P_{in;fi} = 4 \pi (2 \ell + 1) \mid \int_0^{+\infty}
 \phi_{in}(a^{-1} r) j_{\ell}(Q r) R_{n\ell}(r) r^2 dr \mid^2 \label{pril}
\end{eqnarray}
where $n$ and $\ell$ are the appropriate principal and angular quantum
numbers for the final hydrogen-like (${}^4$He$\mu)^{+}$ ion with radial
function $R_{n\ell}(r)$. The choice of the factor $a$ in Eq.(\ref{pril}) is
discussed below. The $j_{\ell}(Q r)$ function is the spherical Bessel
function (see e.g., \cite{LLQ}):
\begin{eqnarray}
 j_{\ell}(x) = \sqrt{\frac{\pi}{2 x}} J_{\ell+ \frac{1}{2}}(x)
\end{eqnarray}
The factor $Q$ is
\begin{eqnarray}
 Q = m_{\mu} v = \sqrt{\frac{2 m_n \Delta E}{(1 + M_4) (1 + m_n + M_4)}} =
 5.825011748
\end{eqnarray}
where $m_n$ = 1838.683662 $m_e$ is the neutron's mass, $\Delta E$ is the
total energy release during the nuclear $(d,t)-$reaction and $M_4 =
7294.2296 m_e$ is the mass of the ${}^4$He nucleus.

In the formulas presented above $\phi_{in}(a^{-1} r)$ is the initial
'post-process' wave function, i.e. the wave function of the system which
arises when the sudden process (i.e. nuclear fusion) is over. The function
$\phi_{in}(a^{-1} r)$ can be found from the bound state wave function $\Psi$
of the initial three-body system. For instance, in the case of nuclear
fusion in the $S(L = 0)-$state of the $dt\mu$ muonic molecular ion one
finds:
\begin{eqnarray}
 \phi_{in}(r_{32}) = \delta_{21} \cdot \Psi(r_{32},r_{31},r_{21}) =
 \Psi(r_{32},r_{32},0) =
 \sum_{i=1}^N C_i \exp(-(\alpha_i + \beta_i) r_{32}) \; \; \; ,
\end{eqnarray}
where $\delta_{21}$ is the nuclear delta-function and $C_i$ are the linear
variational parameters from Eq.(1). These coefficients have been determined
during numerical solution of the Schr\"{o}dinger equation for the initial
three-body system (see Section IV). Note also, that after the 'sudden'
nuclear fusion the new ${}^4$He nucleus arises at the same point '2'. This
does not change the relative $r_{32}$ coordinate which is mass independent.
After the nuclear reaction the $r_{32}$ coordinate becomes the helium-muonic
relative coordinate. In two-body atomic problems, however, it is more
convenient to use the mass-weighted coordinate $r$. The relation between the
relative $r_{32}$ coordinate and mass-weighted $r$ coordinate (which
corresponds to the helium-muonic ion) is written in the form:
\begin{eqnarray}
 r = a r_{32} = \frac{m_{\mu} M_4}{m_{\mu} + M_4} r_{32} \; \; \;,
 \; \; \; or \; \; \; r = \frac{M_4}{1 + M_4} r_{32}
\end{eqnarray}
where in muon-atomic units $m_{\mu} = 1$ and $M_4$ is the nuclear mass (in
muon-atomic units) of the ${}^4$He nucleus. Finally, the initial wave
function $\phi_{in}(a^{-1} r)$ takes the form:
\begin{eqnarray}
 \phi_{in}(a^{-1} r) = \sum_{i=1}^N C_i \exp(-(\alpha_i + \beta_i)
 a^{-1} r) \; \; \; ,
\end{eqnarray}
where $a^{-1} = \frac{1 + M_4}{M_4}$. For the muon and nuclear masses
indicated above, one finds that $a^{-1}({}^4He) = 1.028346555$.

The formulas given above allow us to determine the muon sticking
probabilities for the ground and excited $S(L = 0)$-states in the $dt\mu$
ion. In these calculations we have used our best variational wave functions,
as obtained in this study. After a number of calculations we have evaluated
the total muon sticking probabilities $P_s$ for the ground and excited $S(L
= 0)$-states in the $dt\mu$ ion. These two numerical values are close to
each other. In fact, we have found that each of these two values ($P_s$) is
bounded in the interval 0.008923(3) $\le P_s \le$ 0.008938(3) (for both
bound $S-$states). This means that the total number of $dt-$nuclear
reactions catalyzed by one $\mu^{-}$ muon is $\kappa = P^{-1}_s \approx
112$. This means that before its decay one $\mu^{-}$ muon can catalyze up to
112 nuclear reactions of $dt-$fusion. This number is used in the next
Section to discuss the possibility of exploiting muon catalyzed fusion for
energy production purposes.

\section{Current status of muon-catalyzed nuclear fusion}

Our evaluation of the factor $\kappa = P^{-1}_s \approx 112$ performed above
is based on an assumption that all nuclear $(d,t)-$fusion reactions proceed
from the bound $S-$states of the $dt\mu$ muonic molecular ion. This factor
does not take into account the stripping of muons from rapidly moving He
nuclei and a possibility of nuclear $(d,t)-$fusion in the bound $P-$ and
$D-$states of the $dt\mu$. Formally, an average time required to complete
one $(d,t)-$fusion reaction in any of the $P-$ and/or $D-$states is
substantially longer than the analogous time for the $S-$state(s). In fact,
such a `reaction time' is comparable with the transition time which is
needed to complete all transitions from the bound $P(L = 1), P^{*}(L = 1)$
and $D(L = 2)$ states into the corresponding $S(L = 0)-$state(s). This
directly follows from the fact that actual deuterium-tritium distances are
larger in the $P-$ or $D-$states of the $dt\mu$ ion than in the $S-$states.
Briefly, this means that all $dt\mu$ ions originally formed in the $P-$ and
$D-$states will emit radiation and make transitions into the corresponding
$S(L = 0)-$states. In other words, a possibility of nuclear fusion from the
`rotationally excited' bound $P-$ and $D-$states can be ignored and all
reactions of the nuclear $(d,t)-$fusion in the $dt\mu$ ion can reasonably to
assume to proceed only from the two bound $S-$states of the $dt\mu$ ion. Let
$P$ be exact sticking probability of the muon in the $dt\mu$ ion. The
analogous value $P_s$ is the muon sticking probability determined for the
same $dt\mu$ ion, but only for its bound $S(L = 0)-$states. As follows from
the discussion above we can replace the factor $\kappa = P^{-1}$ by the
approximate value $\kappa \approx P^{-1}_s$.

There are also a few other corrections which can change the numerical value
of the factor $\kappa = P^{-1}_s \approx 112$. The largest of such
corrections corresponds to the `muon stripping' during collisions of the
fast (${}^4$He$\mu)^{+}$ ion with neutral hydrogen molecules. However, even
such a correction cannot change the predicted value of $\kappa$ by 40 \%
\cite{MS}. In other words, the maximal value of $\kappa$ is $\approx$ 160 -
170. On the other hand, to reach break-even, i.e. to compensate for the
energy spent for creation of one $\mu$ muon ($\approx$ 8000 $MeV$ \cite{MS},
\cite{Petr}), one muon needs to catalyze at least 2285 nuclear
$dt-$reactions. In this evaluation we have ignored all possible energy
losses and assumed 100 \% efficiency for each muon. In reality any
thermal-to-electrical conversion has only $\sim$ 30 \% efficiency and only
$\sim$ 70 \% of all muons can produce the maximal number of fusion
reactions. With all these corrections one finds that the factor $\kappa$
must be $\approx$ 8,000 - 11,000 to reach break-even. Such values are
$\approx$ 65 times larger than the maximal value of $\kappa$ which has been
measured experimentally ($\kappa \approx 150$). If somehow in future
experiments the numerical value of $\kappa$ will be increased up to 500,
even then it will be $\approx$ 20 times smaller the value which is needed to
reach break-even. This indicates clearly that muon catalysis of nuclear
reactions cannot be used for energy production purposes.

It should be mentioned that originally the idea to use $\mu^{-}$-muons for
production of repetitive nuclear reactions between light nuclei of hydrogen
isotopes was proposed more than sixty years ago \cite{Frank}. Based on an
obvious chemical analogy these processes were called the muonic catalysis of
nuclear reactions. It was confirmed in \cite{Alv} experimentally by
observing two consecutive $(p,d)-$nuclear reactions catalyzed by the same
muon. The first numerical computations of the bound states in three-body
muonic molecular ions were performed by Belyaev et al in 1959 \cite{Bel} who
found only 20 bound states in six ions $pp\mu, pd\mu, pt\mu, dd\mu, dt\mu$
and $tt\mu$. The overall accuracy of the procedure used in \cite{Bel} was
very low and the authors could not confirm the boundness of the excited
$P^{*}(L = 1)-$states (or (1,1)-states) in the $dd\mu$ and $dt\mu$ ions. It
was concluded only that, if such states are bound, then they are very weakly
bound. The binding energy of these two states was expected to be smaller
than 4.5 $eV$, i.e. smaller than the binding energy of a typical molecule.
Immediately after publication of \cite{Bel} an intense stream of
speculations started about a possible interference (or resonance) between
the formation of excited $P^{*}(L = 1)-$states (or (1,1)-states) in the
$dd\mu$ and $dt\mu$ muonic molecular ions and different atomic/molecular
processes in surrounding molecules (see, e.g., \cite{Zeld} and references
there in). Finally, in a few experimental studies performed by Bystritskii
et al (see \cite{Exp1} and \cite{Exp2} and references therein) it was shown
that one muon can catalyze approximately 10 - 20 $(d,d)-$nuclear reactions
in liquid deuterium (D$_2$) and 90 - 110 $(d,t)-$reactions in the liquid
equimolar deuterium-tritium mixture (D$_2$ : T$_2$ = 1:1). Such very large
numbers of nuclear reactions catalyzed by one muon can be explained only by
the resonance (or very fast) formation of $dd\mu$ and $dt\mu$ muonic
molecular ions. Correspondingly, these processes were called `resonance'
muon-catalyzed fusion of nuclear reactions, in contrast with the `regular'
muon-catalyzed fusion observed in \cite{Alv}.

In experiments performed in 1980's the total number of nuclear reactions
catalyzed by one muon (i.e. the numerical value of the factor $\kappa$
defined above) for the equimolar deuterium-tritium mixture were evaluated a
s $\approx$ 150 (see discussion and references in \cite{MS}). This value is
$\approx$ 15 times smaller than the value which is needed for theoretical
break-even and $\approx$ 65 times smaller than necessary for actual
break-even. Therefore, we have to conclude that all discussion of the
`bright future' for applications of the resonance muon-catalized future for
the energy production purposes appears to be groundless.

\section{Conclusion}

We have considered the problem of highly accurate calculations of bound
states in the three-body muonic molecular ions $pp\mu, pd\mu, pt\mu, dd\mu,
dt\mu$ and $tt\mu$. The study of bound state spectra in the muon-molecular
ions is of interest for solving some theoretical problems and in a number of
applications. In fact, our present knowledge of the bound state spectra in
Coulomb three-body systems with unit charges is essentially based on
knowledge of the spectra of the muonic molecular ions. Note that all muonic
molecular ions can easily be created in real experiments and their various
properties can be measured quite accurately. From a certain point of view,
theoretical and experimental study of these ions is more interesting and
informative than the traditional analysis of atomic three-body (i.e.
two-electron) systems.

The results of variational computations of the total energies for various
bound states in the various muonic molecular ions are presented in Tables I
- VI. Table VI contains the most accurately known predictions for
variational energies in muon-atomic units for each of 22 known bound states
for muonic molecular ions. The accuracy of these total bound state energies
significantly exceeds the accuracy achieved in earlier studies. It should be
mentioned that first variational calculations of muonic molecular ions
started almost 45 years ago \cite{Halp}, \cite{Cart1}, \cite{Cart2},
\cite{DK}. In these works only $S(L = 0)-$ and $P(L = 1)-$states of muonic
molecular ions were considered. It is interesting to note that at that time
the non-variational calculations (see, e.g., \cite{Vin} and references
therein) of muonic molecular ions had a comparable overall accuracy for many
bound $S(L = 0)-$ and $P(L = 1)-$states in muonic molecular ions. In
addition to this, the non-variational methods allow one to determine the
total energies of the bound $D(L = 2)-$ and $F(L = 3)-$states in muonic
molecular ions \cite{Vin}.

Our first variational calculations of muonic molecular ions started 25 years
ago \cite{FrEf}. In particular, the first successful variational
computations of the weakly bound $P^{*}(L = 1)-$states in the $dd\mu$ and
$dt\mu$ ions were performed in our work \cite{FrEf} and also in \cite{Bha1}.
However, at that time we could not compute the bound $D-$ and $F-$states in
the $dd\mu, dt\mu$ and $tt\mu$ ions. Furthermore, our maximal accuracy
achieved at that time was relatively low. Currently, the same energies for
all $S(L = 0)-$ and $P(L = 1)-$states in muonic molecular ions obtained in
\cite{FrEf} can be reproduced with the use of only 20 - 30 exponential basis
functions in Eq.(\ref{e1}), with carefully chosen non-linear parameters
$\alpha_i, \beta_i$ and $\gamma_i$ in each basis function. The first
variational computations of the bound $D(L = 2)-$states in the $dt\mu,
dd\mu$ and $tt\mu$ ions were performed in 1986 \cite{Fro86}, while analogous
calculations of the $F(L = 3)-$state in the $tt\mu$ ion were conducted 15
years later \cite{Fro2001}. The bound $D-$state in the $dt\mu$ ion was also
calculated (variationally) by Kamimura in 1988 \cite{Kamim}.

By using our highly accurate wave functions we have determined the
expectation values of some bound state properties of muonic molecular ions.
We also discuss the problem of mass shifts in muonic molecular ions,
including the case of weakly bound states. A separate (but very important)
problem is to evaluate the muon sticking probabilities for the $S(L =
0)-$states in the $dt\mu$ ion. By using these probabilities we estimated the
total number of the reactions of $(d, t)-$fusion which can be catalized by
one $\mu^{-}$ muon in a liquid equimolar D$_2$:T$_2$ mixture. The current
status of the `resonance' muon-catalyzed nuclear fusion is briefly
discussed. It is shown that this process cannot be used for energy
production purposes.

\begin{center}
    {\bf Acknowledgements}
\end{center}

It is a pleasure to acknowledge the University of Western Ontario for
financial support.

\newpage
  \begin{table}[tbp]
   \caption{The total energies ($E$) of the ground bound $S(L =
            0)-$states in the non-symmetric muonic molecular ions
            in muon-atomic units ($m_{\mu} = 1, \hbar = 1, e = 1$).
            $N$ designates the number of basis functions used in
            Eq.(2).}
     \begin{center}
     \begin{tabular}{llll}
      \hline\hline
$N$ & $E(pd\mu)$ & $E(pt\mu)$ & $E(dt\mu)$ \\
      \hline
 3300 & -0.512 711 796 501 514 087 & -0.519 880 089 782 919 741 & -0.538 594 975 061 413 075 \\

 3500 & -0.512 711 796 501 514 175 & -0.519 880 089 782 919 852 & -0.538 594 975 061 413 084 \\

 3700 & -0.512 711 796 501 514 250 & -0.519 880 089 782 919 605 & -0.538 594 975 061 413 089 \\

 3840 & -0.512 711 796 501 514 283 & -0.519 880 089 782 920 006 & -0.538 594 975 061 413 095 \\
     \hline
  $A$ & -0.512 711 796 501 514 55(1) & -0.519 880 089 782 921 0(5) & -0.538 594 975 061 413 3(2) \\
     \hline
 $E^a$ & -0.512 711 796 501 509 \cite{Fro2001} & -0.519 880 089 782 914 \cite{Fro2001} &
         -0.538 594 975 061 413 \cite{Fro2001} \\

 $E^a$ & -0.512 711 796 500 8 \cite{FSB} & -0.519 880 089 781 9 \cite{FSB} &
-0.538 594 750 58 \cite{FSB} \\
    \hline\hline
  \end{tabular}
  \end{center}
${}^{(a)}$The best variational energies known from earlier calculations. \\
  \end{table}
   \begin{table}[tbp]
    \caption{The total energies ($E$) of the excited bound $S(L =
             0)-$states in the muonic molecular ions in muon-atomic
             units ($m_{\mu} = 1, \hbar = 1, e = 1$). $N$ designates
             the number of basis functions used in Eq.(2).}
      \begin{center}
      \begin{tabular}{llll}
         \hline\hline
$N$ & $E(dd\mu)$ & $E(dt\mu)$ & $E(tt\mu)$ \\
     \hline
 3300 & -0.479 706 380 368 904 996 317 & -0.488 065 357 851 706 559 & -0.496 762 894 249 561 323 055 \\

 3500 & -0.479 706 380 368 904 996 742 & -0.488 065 357 851 706 622 & -0.496 762 894 249 561 323 901 \\

 3700 & -0.479 706 380 368 904 997 024 & -0.488 065 357 851 706 665 & -0.496 762 894 249 561 324 665 \\

 3840 & -0.479 706 380 368 904 997 240 & -0.488 065 357 851 706 683 & -0.496 762 894 249 561 325 040 \\
     \hline
  $A$ & -0.479 706 380 368 904 997 8(2) & -0.488 065 357 851 706 80(5) & -0.496 762 894 249 561 327(1) \\
     \hline
 $E^a$ & -0.479 706 380 368 904 25 \cite{Fro2001} & -0.488 065 357 851 705 \cite{Fro2001} &
         -0.496 762 894 249 561 31 \cite{Fro2001} \\

 $E^a$ & -0.479 706 380 368 \cite{FSB} & -0.488 065 357 841 \cite{FSB} & -0.496 762 894 248 \cite{FSB} \\
      \hline\hline
  \end{tabular}
  \end{center}
${}^{(a)}$The best variational energies known from earlier calculations. \\
  \end{table}
   \begin{table}[tbp]
    \caption{The total energies ($E$) of the bound $S(L = 0)-$states in
             the symmetric muonic molecular ions in muon-atomic units
             ($m_{\mu} = 1, \hbar = 1, e = 1$). $N$ designates the number
             of basis functions used in Eq.(2).}
      \begin{center}
      \begin{tabular}{llll}
        \hline\hline
 $N$ & $E(pp\mu)$ & $E(dd\mu)$ & $E(tt\mu)$ \\
     \hline
 3300 & -0.494386 820248 934808 76255 & -0.531111 135402 386448 65 & -0.546374 225613 816728 844 \\

 3500 & -0.494386 820248 934808 76280 & -0.531111 135402 386449 61 & -0.546374 225613 816728 849 \\

 3700 & -0.494386 820248 934808 76289 & -0.531111 135402 386450 59 & -0.546374 225613 816728 855 \\

 3840 & -0.494386 820248 934808 76294 & -0.531111 135402 386451 22 & -0.546374 225613 816728 856 \\
     \hline
  $A$ & -0.494386 820248 934808 76325(10) & -0.531111 135402 386455(2) & -0.546374 225613 816728 90(3) \\
     \hline
 $E^a$ & -0.494386 820 248 934 546 94 \cite{Fro2001} & -0.531111 135402 386374 5 \cite{Fro2001} &
         -0.546 374 225 613 816 71 \cite{Fro2001} \\

 $E^a$ & -0.494386 820 248 58 \cite{FSB} & -0.531 111 135 402 \cite{FSB} &
 -0.546 374 225 598 \cite{FSB} \\
     \hline\hline
  \end{tabular}
  \end{center}
${}^{(a)}$The best variational energies known from earlier calculations. \\
  \end{table}
   \begin{table}[tbp]
    \caption{The total energies ($E$) of the bound $P(L = 1)-$states in
             the symmetric muonic molecular ions in muon-atomic units
             ($m_{\mu} = 1, \hbar = 1, e = 1$). $N$ designates the number
             of basis functions used in Eq.(2).}
      \begin{center}
      \begin{tabular}{llll}
        \hline\hline
$N$ & $E(pp\mu)$ & $E(dd\mu)$ & $E(tt\mu)$ \\
         \hline
 3300 & -0.468 458 436 303 385 268 11 & -0.513 623 956 792 681 889 & -0.533 263 449 820 383 210 \\

 3500 & -0.468 458 436 303 385 272 62 & -0.513 623 956 792 681 896 & -0.533 263 449 820 383 224 \\

 3700 & -0.468 458 436 303 385 276 22 & -0.513 623 956 792 681 901 & -0.533 263 449 820 383 236 \\

 3840 & -0.468 458 436 303 385 278 58 & -0.513 623 956 792 681 905 & -0.533 263 449 820 383 242 \\
     \hline
  $A$ & -0.468 458 436 303 385 35(5) & -0.513 623 956 792 681 95(2) & -0.533 263 449 820 383 28(2) \\
     \hline
 $E^a$ & -0.468 584 363 033 834 4 \cite{Fro2001} & -0.513 623 956 792 680 25 \cite{Fro2001} &
         -0.533 263 449 820 376 6 \cite{Fro2001} \\

 $E^a$ & -0.468 584 363 033 \cite{FSB} & -0.513 623 956 792 6 \cite{FSB} &
 -0.533 263 449 817 6 \cite{FSB} \\
    \hline\hline
  \end{tabular}
  \end{center}
${}^{(a)}$The best variational energies known from earlier calculations. \\
  \end{table}
   \begin{table}[tbp]
    \caption{The total energies ($E$) of the bound $D(L = 2)-$states of the
             the $dd\mu, tt\mu$ and $dt\mu$ muonic molecular ions in muon
             atomic units. $N$ designates the number of basis functions
             used in Eq.(2).}
      \begin{center}
      \begin{tabular}{lllll}
        \hline\hline
$N$ & $E(dd\mu)$ & $E(tt\mu)$ & $N$ & $E(dt\mu)$ \\
      \hline
3000 & -0.488 708 332 3489 & -0.512 568 653 4551 & 3300 & -0.500 118 083 958 73 \\

3200 & -0.488 708 332 3559 & -0.512 568 653 5939 & 3500 & -0.500 118 083 958 91 \\

3400 & -0.488 708 332 3604 & -0.512 568 653 7061 & 3700 & -0.500 118 083 959 06 \\

3600 & -0.488 708 332 3625 & -0.512 568 653 7938 & 3840 & -0.500 118 083 959 15 \\

3800 & -0.488 708 332 3638 & -0.512 568 653 8630 & 4000 & -0.500 118 083 959 21 \\
     \hline
  $A$ & -0.488 708 332 375(10) & -0.512 568 654 10(11) & $A$ &
        -0.500 118 083 959 35(5) \\
     \hline
 $E^a$ & -0.488 708 332 3645 \cite{Fro2001} & -0.512 568 653 9435
\cite{Fro2001} & $E^a$ & -0.500 118 083 959 201 \cite{Fro2001} \\

 $E^a$ & -0.488 708 331 28 \cite{FSB} &  -0.500 118 083 90 \cite{FSB} &
 $E^a$ & -0.512 568 651 99 \cite{FSB} \\
     \hline\hline
  \end{tabular}
  \end{center}
${}^{(a)}$The best variational energies known from earlier calculations. \\
  \end{table}
\newpage
\begin{table}[tbp]
\caption{The total variational energies in muon atomic units for all
         22 bound states in the six muonic molecular ions.}
  \begin{center}
     \scalebox{0.85}{%
     \begin{tabular}{cccc}
       \hline\hline
ion & $S(L = 0)-$state  & $S^{*}(L = 0)-$state \\
      \hline
$pp\mu$ & -0.494 386 820 248 934 808 763 25(10) & ---------- \\

$pd\mu$ & -0.512 711 796 501 514 55(1) & ---------- \\

$pt\mu$ & -0.519 880 089 782 921 0(5) & ---------- \\

$dd\mu$ & -0.531 111 135 402 386 455(2) & -0.479 706 380 368 904 997 8(2) \\

$dt\mu$ & -0.538 594 975 061 413 3(2) & -0.488 065 357 851 706 80(5) \\

$tt\mu$ & -0.546 374 225 613 816 728 90(3) & -0.496 762 894 249 561 327(1) \\
       \hline
ion & $P(L = 1)-$state  & $P^{*}(L = 1)-$state \\
       \hline
$pp\mu$ & -0.468 458 436 303 385 35(5) & --------- \\

$pd\mu$ & -0.490 664 169 479 327(1) & --------- \\

$pt\mu$ & -0.499 492 029 991 539(1) & --------- \\

$dd\mu$ & -0.513 623 956 792 681 95(2) & -0.473 686 733 842 727 0(5) \\

$dt\mu$ & -0.523 191 456 315 960(1) & -0.481 991 529 973 85(5) \\

$tt\mu$ & -0.533 263 449 820 383 28(2) & -0.489 908 667 504 943 30(8) \\
       \hline
ion & $D(L = 2)-$state  & $F(L = 3)-$state \\
       \hline
$dd\mu$ & -0.488 708 332 375(10)   & --------- \\

$dd\mu$ & -0.500 118 083 959 35(5) & --------- \\

$tt\mu$ & -0.512 568 654 10(11)    & -0.490 554 165 50(25) \\
     \hline\hline
  \end{tabular}}
  \end{center}
  \end{table}
\newpage
   \begin{table}[tbp]
    \caption{The convergence of the $\langle r_{21} \rangle,
             \langle r^2_{31} \rangle, \langle \delta_{21} \rangle$
             and $\nu_{31}$ expectation values for the ground (bound)
             $S(L = 0)-$states of some muonic molecular ions (in
             muon-atomic units).}
      \begin{center}
      \begin{tabular}{lllll}
        \hline\hline
     & $pp\mu$ & $dt\mu$ & $pd\mu$ & $dd\mu$ \\
        \hline
 $N$ & $\langle r_{21} \rangle$ & $\langle r^2_{31} \rangle$ &
 $\langle \delta_{32} \rangle$ & $\nu_{31}$ \\
      \hline
  3300 & 3.299486184357381476285 & 5.8818538929815795 & 0.173456203086 & -0.9466714285 \\

  3500 & 3.299486184357381476271 & 5.8818538929815783 & 0.173456202965 & -0.9466714315 \\

  3700 & 3.299486184357381476272 & 5.8818538929815769 & 0.173456202754 & -0.9466714323 \\

  3840 & 3.299486184357381476267 & 5.8818538929815760 & 0.173456202768 & -0.9466714337 \\
      \hline
  A    & 3.299486184357381476260(20) & 5.8818538929815755(15) & 0.173456202768(15) & -0.9466714337(14) \\
      \hline\hline
   \end{tabular}
   \end{center}
   \end{table}
\newpage
  \begin{table}[tbp]
   \caption{The bound state properties $X$ computed for the ground
            $S(L = 0)-$state and excited $S^{*}(L = 0)-$state in the
            $pd\mu$ and $dt\mu$ muonic molecular ions (in muon-atomic
            units).}
     \begin{center}
     \scalebox{0.85}{%
     \begin{tabular}{cccc}
        \hline\hline
$X$ & $pd\mu (S(L = 0)-$state & $dt\mu (S(L = 0)-$state & $dt\mu (S^{*}(L =
       0)-$state \\
        \hline
$\langle r_{31}^{-1} \rangle$ &
0.6411463715926726(1) &
0.7227000085308685(3) &
0.5146887540974253(3) \\

$\langle r_{32}^{-1} \rangle$ &
0.7533736132562925(1) &
0.7583156106645547(3) &
0.7053752991037914(3) \\

$\langle r_{21}^{-1} \rangle$ &
0.3690963918459403(3) &
0.4038256690725974(5) &
0.2439333374978041(6) \\
       \hline
$\langle r_{31} \rangle$ &
2.451487588757093(2) &
2.117912246542931(3) &
3.933235704904257(3) \\

$\langle r_{32} \rangle$ &
2.087699148755371(2) &
2.023720495653520(3) &
2.738751041467644(3) \\

$\langle r_{21} \rangle$ &
3.100710403484118(3) &
2.747914131666301(5) &
5.161228965846044(5) \\
        \hline
$\langle r^2_{31} \rangle$ &
8.033494173462087(3) &
5.881853892981577(4) &
22.39719316098318(5) \\

$\langle r^2_{21} \rangle$ &
11.082902111235402(4) &
8.2873253005241016(6) &
30.631300336084518(6) \\

$\langle r^3_{32} \rangle$ &
20.6547093169798(2) &
17.4696964524501(3) &
65.2551239488477(3) \\

$\langle r^3_{21} \rangle$ &
42.147963983631(4) &
27.208343544797(5) &
201.45178140015(6) \\
         \hline
$\langle -\frac12 \nabla^2_1 \rangle$ &
0.2806191952124375(7) &
0.3910764762722844(9) &
0.3836327499261653(9) \\

$\langle -\frac12 \nabla^2_2 \rangle$ &
0.3674609039968256(7) &
0.4218837824381256(9) &
0.4476331914219544(9) \\

$\langle -\frac12 \nabla^2_3 \rangle$ &
0.46041133628342744(4) &
0.50069529865293120(5) &
0.50035761171633245(6) \\
     \hline
$\langle \delta({\bf r}_{31}) \rangle$ &
0.11770973301(5) &
0.15452554329(7) &
0.10723299252(7) \\

$\nu_{31}$ &
-0.898787990(3) &
-0.946671420(5) &
-0.946671472(5) \\

$\nu^a_{31}$ &
-0.8987879287819516 &
-0.9466714310522288 &
-0.9466714310522288 \\

$\langle \delta({\bf r}_{32}) \rangle$ &
0.17345620275(5) &
0.17451466503(7) &
0.17875743059(7) \\

$\nu_{32}$ &
-0.9466716379(8) &
-0.9637483436(10) &
-0.9637483713(10) \\

$\nu^a_{32}$ &
-0.9466714310522287 &
-0.9637483334950246 &
-0.9637483334950246 \\

$\langle \delta({\bf r}_{21}) \rangle$ &
1.4616938(4)$\cdot 10^{-5}$ &
8.870847(1)$\cdot 10^{-7}$  &
7.416512(1)$\cdot 10^{-7}$  \\

$\nu_{21}$ &
5.9192044(3) &
10.644434(5) &
10.644863(5) \\

$\nu^a_{21}$ &
5.9191833130515522 &
10.644186704849556 &
10.644186704849556 \\

$\langle \delta_{321} \rangle$ &
2.28129(2)$\cdot 10^{-5}$ &
1.61921(3)$\cdot 10^{-6}$ &
1.35816(3)$\cdot 10^{-6}$ \\
      \hline\hline
  \end{tabular}}
  \end{center}
${}^{(a)}$The exact value.
  \end{table}
\newpage
   \begin{table}[tbp]
    \caption{Calculations of the mass gradients for the $P(L = 1)-$states
             in the $pp\mu, dd\mu$ and $tt\mu$ muonic molecular ions in
             muon-atomic units where $\hbar = 1, e = 1$ and $m_{\mu} =
             206.768262 m_e = 1$. The notations used in this Table
             correspond to the notations used in Eq.(6).}
      \begin{center}
      \begin{tabular}{lll}
        \hline\hline
      & $pp\mu$ & $dd\mu$ \\
        \hline
 $E(m_{\mu} - 2 h)$ & -0.46845 38220 26274 128170 & -0.51361 95148 47688 534232 \\

 $E(m_{\mu} - h)$   & -0.46845 61291 64644 884357 & -0.51362 17147 27907 740353 \\

 $E(m_{\mu} + h)$   & -0.46846 07434 42495 278996 & -0.51362 61777 62223 263542 \\

 $E(m_{\mu} + 2 h)$ & -0.46846 30505 81974 916597 & -0.51362 83987 29794 443654 \\
        \hline
 $\alpha$ & -2.30713 89251 97(10)$\cdot 10^{-3}$ &
            -2.23503 27015 23(10)$\cdot 10^{-3}$ \\
      \hline
      & $tt\mu$ & $(tt\mu)^{*}$ \\
        \hline
 $E(m_{\mu} - 2 h)$ & -0.53325 87089 79317 1418 & -0.48990 43312 78558 0219 \\

 $E(m_{\mu} - h)$   & -0.53326 10794 00655 8570 & -0.48990 64993 92236 3403 \\

 $E(m_{\mu} + h)$   & -0.53326 58202 38499 2057 & -0.48991 08356 16741 7247 \\

 $E(m_{\mu} + 2 h)$ & -0.53326 81906 55003 8459 & -0.48991 30037 27439 5589 \\
        \hline
 $\alpha$ & -2.37041 89216 74(11)$\cdot 10^{-4}$ &
            -2.16811 22634 62(11)$\cdot 10^{-4}$ \\
      \hline\hline
   \end{tabular}
   \end{center}
   \end{table}
\newpage
   \begin{table}[tbp]
    \caption{The total and binding energies of a weakly bound $P^{*}(L =
             1)-$state in the $dd\mu$ muonic molecular ion as the function
             of the muon mass $\tilde{m}_{\mu} = m_{\mu} + y$, where
             $m_{\mu} = 206.768262 m_e$ and $y$ is a non-negative real
             number. All energies have been recalculated to muon-atomic
             units where $\hbar = 1, e = 1$ and $m_{\mu} = 206.768262 m_e
             = 1$. The binding energies $\varepsilon$ are given in
             electronoVolts (eV). $N$ is the total number of basis
             functions used in variational expansion.}
      \begin{center}
      \begin{tabular}{lll}
        \hline\hline
 $y$ & $E(N = 3300)$ & $\varepsilon(N = 3300)$ \\
        \hline
  0  & -0.4736867338427258104 & -1.97498808799965 \\

  1  & -0.4712335469779210548 & -1.68187156159084 \\

  2  & -0.4688058829862271553 & -1.39988399201543 \\

  3  & -0.4664034583422745329 & -1.12949080447425 \\

  4  & -0.4640260218725323738 & -0.87131211085906 \\

  5  & -0.4616733706509872075 & -0.62621871045415 \\

  6  & -0.4593453849322646260 & -0.39554090739252 \\

  7  & -0.4570421278654125672 & -0.18165985634448 \\

 7.9 & -0.4549909488076416492 & -0.00875602540622 \\

7.95 & -0.4548776498295269995 & -0.00002551843452 \\

  8  & -0.4547644307696878137 & +0.00854711729854 \\
        \hline
  $y$ & $E(N = 3840)$ & $\varepsilon(N = 3840)$ \\
         \hline
7.95 & -0.4548776454620138526 & -0.00007865184962 \\
      \hline\hline
   \end{tabular}
   \end{center}
   \end{table}
\newpage
   \begin{table}[tbp]
    \caption{The total and binding energies of a weakly bound $P^{*}(L =
             1)-$state in the $dt\mu$ muonic molecular ion as the function
             of the muon mass $\tilde{m}_{\mu} = m_{\mu} + y$, where
             $m_{\mu} = 206.768262 m_e$ and $y$ is a non-negative real
             number. All energies have been recalculated to muon-atomic
             units where $\hbar = 1, e = 1$ and $m_{\mu} = 206.768262 m_e
             = 1$. The binding energies $\varepsilon$ are given in
             electronoVolts (eV). $N$ is the total number of basis
             functions used in variational expansion.}
      \begin{center}
      \begin{tabular}{lll}
        \hline\hline
 $y$ & $E(N = 3700)$ & $\varepsilon(N = 3700)$ \\
        \hline
  0.0  & -0.48199152997371597 & -0.66033868534010 \\

  1.0  & -0.48184465428112405 & -0.31071047997243 \\

  1.8  & -0.48173196328443396 & -0.05573790640587 \\

  1.9  & -0.48171831993605128 & -0.02623616728477 \\

  1.95 & -0.48171161233515334 & -0.01212193523051 \\

  1.98 & -0.48170758889627615 & -0.00365631831137 \\

  2.0  & -0.48170482239836190 & +0.00246722246561 \\
        \hline
  $y$  & $E(N = 3840)$ & $\varepsilon(N = 3840)$ \\
         \hline
  1.95 & -0.48171161309408528 & -0.01212624559609 \\

  1.98 & -0.48170759005686002 & -0.00366291081502 \\

  2.0  & -0.48170492020870050 & +0.00191157404752 \\
      \hline\hline
   \end{tabular}
   \end{center}
   \end{table}
\end{document}